\documentclass[doublespacing]{elsart}
\usepackage{graphicx}
\usepackage{amssymb}
\usepackage{amsmath}
\usepackage{amsbsy}
\usepackage{latexsym}
\font\mybb=msbm10 at 10pt
\def\bb#1{\hbox{\mybb#1}}
\newcommand{\IG}{\includegraphics}

\usepackage{color}
\usepackage{colortbl}
\newcommand{\roig}[1]{\textcolor[rgb]{1,0,0}{#1}}           
\newcommand{\verd}[1]{\textcolor[rgb]{0,1,0}{#1}}           
\newcommand{\blau}[1]{\textcolor[rgb]{0,0,1}{#1}}           

\begin{document}
\begin{frontmatter}

\title{Evaluation of modified Hantush's function}
\vspace{-0.8 cm}
\author{\corauthref{cor1} Tatyana V. Bandos},
\author{Koldobika Martin-Escudero,}
\author{Jos\'e M. Sala-Lizarraga}
\vspace{-0.6 cm}

\address{Grupo de Investigaci\'on ENEDI, Dpto. de Ingenier\'ia Energ\'etica
Universidad  del Pa\'is  Vasco, Alda. Urquijo s/n, 48013 Bilbao, Spain}
\vspace{+0.6 cm}

\corauth[cor1]{Corresponding author. Tel: +34 946018249; Fax: +34 946014040. \\
               E-mail address: tbandos@uv.es (T. V. Bandos).}

\begin{abstract}
It is pointed out that Hantush's \emph{M} function, commonly used in groundwater pumping modeling, is identical to the  function known
in the problems of heat conduction in the ground.
 A modified Hantush  function  $E$   used in the steady-state solutions  for heat  conduction that take into account advection due to groundwater flow around borehole heat exchangers (BHEs) is introduced.
New exact representation and   two-types of approximate  expressions for this two-parametric function \emph{E} are  presented:
one is suitable for small values of a parameter  and  the other for its large values.
 High accuracy of the approximate formulae is verified by comparison
with the exact values provided by modified Hantush's function $E$ in those validity domains.
\\
These analytical expressions can be used for designing complex  borehole  configurations and may be potentially  useful
in steady-state analysis of   partially penetrated wells (PPWs) in leaky aquifer.
\end{abstract}
\begin{keyword}
Hantush's  Well Function;  Leaky Well Function; Generalized Incomplete Gamma Function;\\
Partially Penetrating Wells;  Borehole Heat Exchangers; Groundwater;
\end{keyword}
\end{frontmatter}
\vspace{-1.3 cm}\textbf{}
\section*{Highlights}
\begin{itemize}
\item
 Modified Hantush's integral $E$ is approximated in a wide range of two-parameter space.
 \item
 The  maximum relative error of the  low-order asymptotic expansions is less than $0.01\%$  with respect to numerically calculated  modified Hantush's integral.
 \item
 Analytical  form of  $E$ shows  approaches to  Hantush's  well functions.
 \end{itemize}
 
\setcounter{page}{1}
\newpage
\section*{Introduction}
\label{sec:pd}
In this study we will address the problem of analytical approximation for the integral  $E$ widely used in geohydrology and geothermics.
These  may be useful to decrease computational costs of designing wells in leaky aquifers and borehole heat exchangers with advection. 
\\
Integral that arises in the solutions for  heat conduction models without taking into account the advection is 
\begin{eqnarray}
\label{eqn:M}
&&\emph{M}(u,h) =2\cdot\int\limits_{\sqrt{u}}^\infty erf(hv) \exp(-v^2)dv/v, \; \; u\geq0, h\geq0,
\end{eqnarray}
 where $erf$ is the error function. 
 It is identical to the Hantush  function   $M(u,h)$ known in the geohydrologic literature  (Hantush, 1961).  This analogy between transient solutions  for drawdown and temperature  that account for axial variations  by parameter $h$ in  Eq.(1)  is lacking,  to the best of our knowledge,   in the literature on geohydrology  and geothermics.\\
 In modeling of well hydraulics,  the parameters of the integral (\ref{eqn:M}), $u$, which is proportional to inverse time $t$,  and $h$  correspond to the aquifer properties in the three-dimensional  description  of  unsteady drawdown  distribution   around  partially penetrating wells   (Hantush, 1964).
 Hantush's function   can be evaluated  by numerical integration or using some series expansion for   $M(u,h)$
obtained in a number of papers   (Barry et al., 1999; Hantush, 1961; Harris, 2008;  Trefry, 1998; Trefry, 2005; Veling and Maas, 2010; Yeh  and Chang, 2013).
The advantage of approximate expressions for $M(u,h)$ is that they allow physical interpretations (Trefry  and Townley, 1998), while  the results of its numerical study are often not suggestive,  particularly for very large values of  $u$ and $h$.
 Asymptotic  and Maclaurin series  for the integral (1) were derived  to evaluate unsteady flow around PPWs  with  accurate analytical formulae as alternative  to calculation by numerical quadrature (Trefry, 2005).
The approximate expressions to the integral in  Eq.(\ref{eqn:M}) have been obtained and further developed (Bandos et al., 2009;  Bandos et al., 2011) for $u \ll 1  \ll u\cdot h^2$
and  $u\cdot h^2\ll 1$    when approaching steady-state limit  $u\rightarrow 0 $ (i.e. $t \rightarrow \infty$).  These approximate formulae have been successfully applied to hydrological and geothermal studies   (Jia et al., 2022; Yeh  and Chang, 2013).
\\
The present study is devoted to  derivation of new   exact and approximate representation of the following integral:
\begin{eqnarray}
\label{eqn:A4}
&& \emph{E}(b, h) =\int\limits_0^\infty erf(hv) \exp(-v^2)\cdot exp(-b/v^2)dv/v, \; \; b\geq0, h\geq0
\end{eqnarray}
We call $\emph{E}(b, h)$ the modified Hantush function, which reflects  the fact that it  recovers   steady-state  Hantush's function  $M(u\rightarrow 0, h)/2$ in the limiting case of $b=0$.   To describe leakage the parameter $b$ was added to  the groundwater flow equation  for a leaky  aquifer (Hantush and Jacob, 1955; Hantush, 1957) as a volumetric sink/source.

 The function $E(b,h)$ has applications in a wide range of areas from statistics/probability theory to ground-coupled heat pumping, and contaminant  hydrogeology  (Chaudhry and Zubair, 1994; Chiasson and O'Connell, 2011; Claesson and  Hellstr\"om, 2000; Molina-Giraldo et al, 2011; Park and Lee, 2021; Van de Ven et al., 2021; Jia et al., 2022).
In the case of  geothermal heat pump systems,  integral  (\ref{eqn:A4}) arises in the steady-state solution for the 3D temperature  distribution
around   the BHEs with groundwater advection  described by the parameter $b$ (Carslaw  and  Jaeger, 1959; Eskilson, 1987; Molina-Giraldo et al, 2011; Sutton et al., 2003; Zubair and Chaudhry, 1998).
In geohydrology, modified Hantush's function $\emph{E}(b, h)$  appears in the solution for steady axially symmetric drawdown  distribution
around well partially penetrating leaky aquifer (Hantush, 1957; Hantush, 1964; Park and Lee, 2021).\\
\\
The modified Hantush integral $E(b,h)$  is also  mathematically  related  to  well functions used in geohydrology.
First, it is worth to note that in the limiting case  of  $b \rightarrow 0$ in (\ref{eqn:A4}) and $u\rightarrow 0$ in (\ref{eqn:M}) $E(b=0,h)= M(0,h)/2=\sinh^{-1}{h}$. In the other limit  $h\rightarrow\infty$ in (\ref{eqn:A4}),  $E(b,\infty)=K_0(\sqrt{4 b})$, where  $K_0(\beta)$ is the modified Bessel function  of  the second kind of zero order. Therefore,   $E(\beta^2/4,\infty)$ coincides  with  limit  $W(u\rightarrow 0,\beta)/2$ of \emph{well function for leaky aquifers}  $W(u,\beta)$  defined by the following infinite integral (Hantush  and Jacob, 1955)
\begin{eqnarray}
\label{eqn:W}
&& \emph{W}(u,\beta) =2\int\limits_{\sqrt{u}}^\infty \exp(-v^2-{\beta^2\over 4v^2})dv/v, \; \; u\geq0,
\end{eqnarray}
where $\beta$ is real.
In the limiting case  of   $u \rightarrow 0$ ($t\rightarrow \infty$), this leaky aquifer function presents the  radial  steady drawdown distribution
in leaky systems without storage in semipervious layer (Hantush  and Jacob, 1955). 
\\
Second,  in the limiting case of $h\rightarrow\infty$, Hantush's function tends to
the exponential integral, $Ei(-u)$,  $M(u,\infty)=-Ei(-u)$.  Furthermore, in the  case of $\beta\rightarrow 0$, the function (\ref{eqn:W}) tends to the same limit  as  $M(u,\infty)$ and  becomes
the \emph{well function for non-leaky aquifers} $W(u)$,  $W(u,0)=-Ei(-u)=W(u)$.
This integral solution  $W(u)$ introduced by  Theis (1935), using   analogy between  Darcy's law and Fourier's law within the context of  well hydraulics,    describes  a nonsteady radial groundwater  flow  around well (Hantush, 1964).
In geothermics,  the integral $W(u)$   is known as solution to  Kelvin's infinite line-source of heat model, while solution to the so-called  moving  infinite line-source (MILS) model  described by Carslaw and Jaeger (1959) is based on the integral $W(u,\beta)$.
 Time-series  expansion for  the $\emph{W}(u,\beta)$  function  (Hantush  and Jacob, 1955; Hunt, 1977; Veling and Maas, 2010), which have been successfully applied to  geohydrological problems,  can be  adapted in the transient temperature solutions that account for advection and are in focus of  experimental and  numerical investigations (Van de Ven  et al., 2021; Barbieri et al., 2022).
\\
 Approximate  expressions for both  well functions  $M(u, h)$  and $W(u,\sqrt{4 b})$ (Hantush and Jacob,  1955; Hunt, 1977; Trefry, 2005), which  appear as  limits of the  integral $E(b, h)$  (i.e. $E(b, h\rightarrow\infty)=W(0,\sqrt{4 b})/2$ and $E(b \rightarrow0, h)=M(0,h)/2$ ) are widely used and revisited (Barbieri et al., 2022; Pasquier, P. and L. Lamarche, 2022; Veling and Maas, 2010). Therefore,  deriving analytical formula  for modified Hantush's $E(b, h)$  that take  into account  both leakage  and  axial effects described by parameters $b$ and $h$  is desirable even under steady-state conditions.
 A  literature  search reveals no approximations  for  the  3D steady-state solution $E(b, h)$ in geohydrology,  while in geothermics the approximate solutions  for temperature  are available for small and large values of the  parameter $bh^2$, but only at a fixed  parameter $b$   (Eskilson, 1987).
\\
This paper presents (i) explicit expression for the $E(b, h)$  integral  in terms of  power series and special functions; (ii) analytical formulae for its asymptotic behavior   at one segment   $b\cdot h^2\geq 1\geq b$ of  the $\{b,h\}$ parameter space, and (iii)
approximate expressions for this  function for $b<b\cdot h^2\leq 1$  at other  segment.
\\
The rest of the paper is organized as follows.
Section \ref{sec:E}  proposes new  approximations for the  integral (\ref{eqn:A4}) at  two segments of the parameter  domain  (in  Sections \ref{sec:MILS} and \ref{sec:MFLS})  and verifies  these results,   comparing them with  the predictions  known in the  fields of  geothermics  and geohydrology.
The findings are then analyzed  and  validated  numerically  in Section \ref{sec:PRE}.
Section  \ref{sec:con}  contains conclusions  of  the paper.
\section{Evaluation of integral \emph{E(b,h)}}\label{sec:E}
 \label{eqn:FS}
To properly analyze the finite-length  effects in the heat exchange with the groundwater flow, the function  $E(b, h\rightarrow\infty)=W(0,\sqrt{4b})/2$
is subtracted from the integral
$E(b,h)$, by  expressing    $erf(x)=1-erfc(x)$  in terms of the  complementary error function $erfc(x)$.  Substituting  this in the  integral  Eq.(\ref{eqn:A4}) we obtain
\begin{eqnarray}
\label{eqn:B}
 &&E =  K_0(\sqrt{4\emph{b}})-A(b, h),\nonumber\\
 \\
 &&K_0(\sqrt{4\emph{b}})=\int\limits_0^\infty  \exp(-u^2-b/u^2)du/u\nonumber
\end{eqnarray}
\begin{eqnarray}
\label{eqn:A}
 A(b, h)=\int\limits_0^\infty erfc(hu) \exp(-u^2-b/u^2)du/u
\end{eqnarray}
Note that,  in the limiting case of $h\rightarrow\infty$ the integral (\ref{eqn:A}) becomes zero recovering in (\ref{eqn:B}) the steady-state solution  for the MILS model or the steady-state solution for leaky aquifer (Hantush  and Jacob, 1955).   As it is shown below $A(b, h)$ can be expressed explicitly  as power series.\\
Here we  will establish  approximations for $A(b, h)$ in (\ref{eqn:A})  by expanding  the exponential function
in the integrand of Eq.({\ref{eqn:A4}) following Hantush's approach (Hantush  and Jacob, 1955; Hunt, 1977).
After a suitable change of the integration variable in Eq.({\ref{eqn:A})  (i.e. $hu\rightarrow u$)  the exponential function  in the  integrand  of
$A$   is replaced by its convergent power series in $1/h^2$,\\
\begin{eqnarray}\textbf{}
\label{eqn:B03}
exp(-u^2/h^2)=\sum\limits_{m=0}^\infty(-u^2/h^2)^m/m!
\end{eqnarray}
then,   the integral of Eq.(\ref{eqn:A}) is evaluated term by term (Hantush  and Jacob, 1955) with the
use of the integral representation  for  the exponential integral  (Gradshteyn  and   Ryzhik, 1971)
\begin{eqnarray}\label{eqn:Ei}
&&\int\limits_0^\infty erfc(hu)exp(-b/u^2)du/u=-Ei(-h\sqrt{4 b})
\end{eqnarray}
This equation can be equivalently  re-written  as
\begin{eqnarray}\label{eqn:Ei_0}
&&\int\limits_0^\infty{1\over \sqrt{\pi}} \Gamma(1/2,h^2u^2)exp(-b/u^2)du/u= \Gamma(0,h\sqrt{4 b})
\end{eqnarray}
where $\Gamma(\alpha, x)$ is the incomplete Gamma function (Abramovitz and Stegun, 1965) and  $\Gamma(1/2,x^2)=\sqrt{\pi}erfc(x)$  (Prudnikov et al., 1983).
In such a way, we find
\begin{eqnarray}
\label{eqn:AG}
&&A(b, b_H)=\int\limits_0^\infty erfc(u) \exp(-u^2b/b_H-b_H/u^2)du/u\\
&&=-Ei(-\sqrt{4 b_H})I_0(\sqrt{4 b}) \textbf{+}{\bb S}(b,b_H), \qquad\nonumber\\
\nonumber\\
&&b_H=h^2\cdot b
\end{eqnarray}
where    $I_0 (\sqrt{4 b})$   is the modified Bessel function of the first kind and zero order (Prudnikov et al., 1983),
\begin{eqnarray}\label{eqn:I0}
I_0 (\sqrt{4 b})=\sum\limits_{m=0}^\infty{b^m \over m!m!}
\end{eqnarray}
We are going to prove (\ref{eqn:AG}) in which the  term {\bb S}  can  be expressed explicitly as follows
\begin{eqnarray}
\label{eqn:S}
&&{\bb S}(b,b_H)= \sum\limits_{m=1}^\infty{(-b/b_H)^m\over m!}\cdot{1\over 2\sqrt{\pi}}\sum\limits_{f=0}^{m-1}(-b_H)^{ f}(m-f-1)!/m!\\
&&\times \Gamma(m+1/2-f,0;b_H),\qquad \nonumber
\end{eqnarray}
where $\Gamma(\alpha,0;b)$ is the so-called the Kr\"atzel function (Kilbas et al, 2006)
\begin{eqnarray}
\label{eqn:Kr}
&&\Gamma(\alpha,0;b)= 2\int\limits_0^\infty u^{2\alpha}\exp(-u^2-b/u^2)du/u
\end{eqnarray}
appearing as the limiting case of the generalized incomplete gamma function  introduced  by  Chaudhry  and Zubair  (1994),
\begin{eqnarray}
\label{eqn:GG}
&&\Gamma(\alpha,x;b)= 2\int\limits_{\sqrt{x}}^\infty u^{2\alpha}\exp(-u^2-b/u^2)du/u.
\end{eqnarray}
The asymptotic behavior of the Kr\"atzel function $\Gamma(\alpha,0;b)$ at zero and infinity
can be applied to evaluation of integrals (Kilbas et al., 2010)  including the discrete form in (\ref{eqn:S}), which
we will use in sections \ref{sec:MILS}  and \ref{sec:MFLS}}.\\
It is worth to note that  $\Gamma(\alpha,x;b=0)=\Gamma(\alpha,x)$,
\begin{eqnarray}
\label{eqn:GK}
\Gamma(\alpha,0;b)= 2 b^{\alpha/2}K_\alpha(\sqrt{4\emph{b}}),
\end{eqnarray}
and
\begin{eqnarray}
\label{eqn:GW}
\Gamma(0,x;b)=W(x, \sqrt{4\emph{b}}), \hspace{0.3cm}   \Gamma(0,0;b)=W(\sqrt{4\emph{b}}).
\end{eqnarray}
Using   (\ref{eqn:GK})   ${\bb S}$  from Eq.(\ref{eqn:S}) can  be re-written  in terms of the  modified Bessel function of  the third kind  or  MacDonald function (Abramovitz and Stegun, 1965) as follows
\begin{eqnarray}
\label{eqn:AK}
&&
{\bb S}(b, b_H)=\sum\limits_{m=1}^\infty{(-b/b_H)^m\over m!}\cdot{1\over 2\sqrt{\pi}}\sum\limits_{f=0}^{m-1}(-b_H)^{ f}{(m-f-1)!\over m!}\\
&&\times 2b_H^{(m+1/2-f)/2}K_{m+1/2-f}(\sqrt{4 b_H})\qquad\nonumber
\end{eqnarray}
These latter  for integer  $m$ can be expressed  as  a finite sum  (Gradshteyn  and   Ryzhik, 1971)
\begin{eqnarray}
\label{eqn:macD}
&&K_{m+{1\over 2}}(x)=\sqrt{\pi \over 2x }\exp(-x)\sum \limits_{k=0}^m{(m+k)!\over k!(m-k)!(2x)^k},~for  \hspace{0.3cm} ~ m=0, 1, 2,\cdots
\end{eqnarray}\\
Substituting  (\ref{eqn:B03}) in  (\ref{eqn:A})  and integrating term-by-term we obtain the  following power series in $1/h^2$
 \begin{eqnarray}\label{eqn:Asum}
 A(b, b_H)=\sum\limits_{m=\textbf{0}}^\infty(b/b_H)^m{(-1)^m\over m!}\cdot {\bb I}_m(b_H)
 \end{eqnarray}
 where
\begin{eqnarray}\label{eqn:Im}
&&{\bb I}_m(h\sqrt{4\emph{b}}) =\int\limits_0^\infty u^{2m} \cdot erfc(hu)exp(-b/u^2)du/u.\\\nonumber
\end{eqnarray}
As a by-product, integrating ({\ref{eqn:Im}) by the same method we
extend the  list of  known  integrals ${\bb I}_m$ from  $m=0$ (\ref{eqn:Ei}), and  $m=1$ (Gradshtein and Ryzhik, 1967; Prudnikov et al, 1983)
to an arbitrary $m$
 \begin{eqnarray}\label{eqn:Eim}
&&{\bb I}_m(c)=-Ei(-\sqrt{c}) (-c/4)^m/m! ~ \hspace{1.5cm}  \\
&&+{1\over m!}\cdot{1\over\sqrt{\pi} }\sum\limits_{f=0}^{m-1}(-1)^f (m-f-1)! \cdot (c/4)^{(m+1/2+f)/2}K_{m+1/2-f}(\sqrt{c})\nonumber\\
&&\hspace{7.8cm} for  \hspace{0.8cm} ~ m=0,1, 2\cdots\nonumber
\end{eqnarray}
In particular, for the case $m=2$ the   proposed  formula   gives
\begin{equation}
 \label{eqn:Im2}
{\bb I_2(c)=-Ei(-\sqrt{c}) \big({c/ 4}\big)^2/2!
 + \big(K_{5/2}(\sqrt{c})-\sqrt{c/4}K_{3/2}(\sqrt{c})\big)\cdot{({c/4})^{5/4} \over 2!\sqrt{\pi}}}
\end{equation}
which can be expressed in elementary functions.
The finite series ({\ref{eqn:Eim}) for the integral ${\bb I}_m$, which are not found in (Prudnikov et al., 1983),  can be verified  for any integer $m$ by using symbolic techniques (Wolfram, 2003).\\
Therefore, we have proved Eqs.(\ref{eqn:AG})   with the double sum ${\bb S}$  in the form of (\ref{eqn:S})  by  substituting the expression  (\ref{eqn:Eim})   in Eq.(\ref{eqn:Asum}) and  identifying the series (\ref{eqn:I0}).\\
Furthermore,  substituting $A(b, h)$ (\ref{eqn:AG}) to Eq.(\ref{eqn:B}), we obtain  the power series representation for modified Hantush's function as follows
\begin{eqnarray}
\label{eqn:F}
&&E(b,h) = K_0(\sqrt{4\emph{b}})+Ei(-\sqrt{4 h^2b})I_0(\sqrt{4 b})-{\bb S}(b, h), \nonumber \\  \\
&&{\bb S}=\sum\limits_{m=1}^\infty{(-1/h^2)^m \over [m!]^2}\cdot{1\over 2\sqrt{\pi}}\sum\limits_{f=0}^{m-1}(-h^2b)^{ f}(m-f-1)!\cdot \Gamma(m+{1\over 2}-f, 0; h^2b).\nonumber
\end{eqnarray}
In the following,  the approximate expressions are  derived for  large values of $b\cdot h^2 \geq 1\geq b$, and then for   $b\cdot h^2 \leq 1<h$ in  section \ref{sec:MFLS}. Such ranges  of  parameters $b, h$ are of interest  in the problems of groundwater flow in aquifers and BHEs.

\subsection{Series representation for E(b,h)}
\label{sec:MILS}
To   find    approximate expressions for sufficiently large values of   $b\cdot h^2 \geq 1$   (in the next  section)
the double  sum ${\bb S}(b, h)$ from  Eq.(\ref{eqn:F}) can be decomposed on the sums over  the first column in  $\{m, f\}$ index plane and diagonal $(m=f)$  following Hantush's approach  (Hantush  and Jacob, 1955).
By changing the variable:  $m\rightarrow m+1$  the series ${\bb S}$ in  (\ref{eqn:S}) can be re-written  as
\begin{equation}\label{eqn:B4}
{\bb S}(b_H,h)=\sum\limits_{m=0}^\infty\sum\limits_{f=0}^{m} {(m-f)!(-1)^{m+f+1}\over [(m+1)!]^2} {b_H^{f}\over h^{2(m+1)}}\cdot\Gamma(m+{3\over 2}-f,0;b_H)\cdot{1\over  2\sqrt{\pi}}\qquad
\end{equation}
The series is not computationally expensive, as it converges rapidly due to the  asymptotic properties of the $\Gamma(m+{3\over 2}-f,0;b_H)$ function at infinity and zero.
Extracting the  sums of the terms in the  first column $f=0$, and at the   diagonal $f=m$    results  in
\begin{eqnarray}
\label{eqn:B5}
&& {\bb S}(b_H,h)=\sum\limits_{m=0}^\infty{m!(-1)^{m+1}\over (m+1)!^2} {1\over h^{2(m+1)}}\cdot\Gamma(m+3/2,0;b_H)\cdot{1\over  2\sqrt{\pi}}\qquad\\
&&-\sum\limits_{m=1}^\infty{1\over [(m+1)!]^2} {b_H^{m}\over h^{2(m+1)}}\cdot\Gamma(3/2,0;b_H)\cdot{1\over  2\sqrt{\pi}}\nonumber \\
&&+\sum\limits_{m=1}^\infty\sum\limits_{f=1}^{m} {(m+1-f)!(-1)^{m+f+2}\over [(m+2)!]^2} {b_H^{f}\over h^{2(m+2)}}\cdot\Gamma(m+1+3/2-f,0;b_H)\cdot{1\over  2\sqrt{\pi}}\qquad\nonumber
\end{eqnarray}
where  the index of summation is already changed, $m \rightarrow m-1$ (from $m=2$  to $m=1$), in the third term. Therefore, Eq.(\ref{eqn:B5}) represents a formally  exact series for  {\bb S}  of Eq.(\ref{eqn:AG}). The
remainder  $R(q, p)$, given by the third line in  (\ref{eqn:B5}),  is an infinite series  each term of which involves a finite summation: it can be also written as follows
\begin{eqnarray}\label{eqn:Rq}
&&R(q, p)=\sum\limits_{m=q}^\infty\sum\limits_{f=p}^{m} {(m+1-f)!(-1)^{m+f+2}\over (m+2)!^2} {b_H^{f}\over h^{2(m+2)}}\cdot b_H^{(m+5/2-f)/2}\times \nonumber \\ &&K_{m+5/2-f}(\sqrt{4b_H})\cdot{1\over  \sqrt{\pi}}.
\end{eqnarray}
Its convergence  properties  are expected to be reasonable since the terms, including the powers of  $1/h^2<1$, in the summation over $m$ are also  suppressed by factorial in the denominator.\\
The second   series  in (\ref{eqn:B5}) can be calculated exactly:   using  (\ref{eqn:GK}) and (\ref{eqn:macD}),  and  the identifying  (\ref{eqn:I0})
we reduce it to
\begin{eqnarray}\label{eqn:B7}
&&+{1\over h^2}\big(-1 +(I_0(\sqrt{4 b})-1)/b\big)\cdot b_H^{3/4}K_{3/2}(\sqrt{4b_H})\cdot{1\over  \sqrt{\pi}} \\
&&={\sqrt{b}\over h}\big[-1 +(I_0(\sqrt{4 b})-1)/b\big]\cdot\exp(-\sqrt{4 b_H})[1+ 1/\sqrt{4 b_H}]/2.\nonumber
\end{eqnarray}
 Then, the exact expression (\ref{eqn:F})  can be re-written as 
\begin{eqnarray}\label{eqn:B3}
&&E(b,h) = K_0(\sqrt{4\emph{b}})+Ei(-\sqrt{4 h^2b})I_0(\sqrt{4 b})\nonumber\\ \nonumber\\
&&-\sum\limits_{m=0}^\infty{m!(-1)^{m+1}\over [(m+1)!]^2} {1\over h^{2(m+1)}}\cdot\Gamma(m+3/2,0;4 h^2b)\cdot{1\over  2\sqrt{\pi}}\\
&&+{\sqrt{b}\over h}\big[-1 +(I_0(\sqrt{4 b})-1)/b\big]\cdot\exp(-\sqrt{4 h^2b})[1+ 1/\sqrt{4 h^2b}]/2\nonumber\\
&&-R(q=1,p=1) \nonumber
\end{eqnarray}
and this is the  form suitable  for computations with large $h$ as well as for comparison with the\emph{ well functions for aquifers}  (\ref{eqn:W}):
\begin{eqnarray}
\label{eqn:B3W}
&&E(b,h) = W(0,\sqrt{4\emph{b}})/2-W(\sqrt{4 h^2b},0)I_0(\sqrt{4 b})\nonumber\\ \nonumber\\
&&-{\bb S}(b,h).
\end{eqnarray}
In the following, we will  find approximations for power series Eq.(\ref{eqn:B3W}) over the relevant $\{b, h\}$  parameter space.
\subsubsection{Asymptotic series: approximate expressions for $b\cdot h^2>1$}
\label{sec:asymMFLS}
New approximate expressions for the integral  $E$  (\ref{eqn:B3W})  for small values of  parameter $b\leq1$  and large values of   $b_H=b\cdot h^2>1$   are derived  in the following way.\\
 Use of the   asymptotic
approximation of the $K_{m+3/2}(x) \sim  K_{1/2}(x)[1+{(m+2)!\over m!2x}+ O({1\over x^2})]$  for  $x>>1$  from Eq.(\ref{eqn:macD}) results in a simpler  expression for  the infinite series  ${\bb S}$ (\ref{eqn:B5})  over Kr\"atzel  functions  (\ref{eqn:GK}).
After summing  over  $m$ the series  in powers of $1/h^2$, the first term  in the sum (\ref{eqn:B5}) can be calculated exactly:
\begin{eqnarray}
\label{eqn:B6}
&&\textbf{-}\sum\limits_{m=0}^\infty{m!(-1)^{m+1}\over [(m+1)!]^2}{b^{(m+1)/2}\over h^{(m+1)}}\cdot b_H^{1/4}K_{3/2+m}(\sqrt{4b_H})\cdot{1\over  \sqrt{\pi}} \\
&&=\big(\gamma+\ln\big({\sqrt{b}\over h}\big)-Ei\big(-{\sqrt{b}\over h}\big)+{1\over 4\sqrt{b_H}}[1+\e^{-{\sqrt{b}\over h}}({\sqrt{b}\over h}-1)]\big)\times{\exp( -\sqrt{4b_H})\over 2}\nonumber
\end{eqnarray}
Summation of the above expression and (\ref{eqn:B7}) (giving the first and the second terms in (\ref{eqn:B5}))
results  in
\begin{eqnarray}\label{eqn:B55}
&&\textbf{-}{\bb S}(b, h) = \big(\gamma+\ln\big({\sqrt{b}\over h}\big)+W\big({\sqrt{b}\over h}\big)
+{1\over 4h \sqrt{b}}[1+e^{-{\sqrt{b}\over h}}({\sqrt{b}\over h}-1)]\\ \nonumber\\ \nonumber
&&+{\sqrt{b}\over h}\big[-1 +(I_0(\sqrt{4 b})-1)/b\big]\big[1+{1\over 2h \sqrt{b}}\big]\big)\times{\exp(-2h\sqrt{b})\over 2}\\ \nonumber\\
&&- R(q=1, p=1)\nonumber
\end{eqnarray}
The remainder   $R$ (\ref{eqn:Rq}) can be omitted in Eq.(\ref{eqn:B55})  because of the rapid convergence of its  series for values of  $h^2>>1$ and
$b\cdot h^2>1$.
Retaining the first few terms of the  series in (\ref{eqn:B5}) (or expanding  (\ref{eqn:B55}) in terms of the small parameters  ${\sqrt{b}/ h}$  and $b$),  the approximate expression for $E(b, h)$
from Eq.(\ref{eqn:B3W}),   $\tilde{E}(b,h)$, may be written  as
\begin{eqnarray}\label{eqn:B8}
&&\tilde{E}(b, h)  =  W(0,\sqrt{4\emph{b}})/2-W(h\sqrt{4\emph{b}},0)I_0(\sqrt{4 b})\\
&&+\sqrt{b}/ h\big[ 1-{\sqrt{b}\over 8h} + {b\over 4} + {1\over 2h \sqrt{b}}-{3\over 8h^2} \big]
\times{\exp(-2h\sqrt{b})\over 2}
\nonumber
\end{eqnarray}
where the second term  is exponentially small when $u\rightarrow\infty$  and $b\leq1$, because  $W(u,0)$  asymptotically approaches to
$Ei(-u)\approx -\exp(-u)/u$.\\
Further simplifying  $\tilde{E}$ from (\ref{eqn:B8}) at the first approximation  in  $\sqrt{b}/ h$ gives
\begin{eqnarray}\label{eqn:AB9}
&&\tilde{E}(b,h)  \simeq  W(0,\sqrt{4\emph{b}})/2-W(h\sqrt{4\emph{b}},0)I_0(\sqrt{4 b})\\
&&+\sqrt{b}/ h\big[1+{1\over 2h \sqrt{b}}\big]\times{\exp(-2h\sqrt{b})\over 2}
\nonumber
\end{eqnarray}
Note that Eq.(\ref{eqn:AB9}) involves only the first term $m=0$ of the infinite series (\ref{eqn:B4}), which is rapidly convergent
for $b\cdot h^2\geq 1 \gg b$ due to the asymptotic properties of the $\Gamma(m+3/2,0;b\cdot h^2)$ function at infinity (Kilbas et al., 2010).
The two proposed expressions:  (\ref{eqn:B8}) and  (\ref{eqn:AB9}) are much simpler and sufficient for accurate approximation of the integral $E$ as shown in Section 2.
\\
In the limiting case of $h\rightarrow \infty$  these approximate expressions   for $E(b, h\rightarrow\infty)$ reduce to  the steady-state solution for  the groundwater flow in the infinite  leaky aquifer
(Hantush  and Jacob, 1955), $W(0,\sqrt{4b})/2$,
and  are consistent with the prediction  of the  moving infinite line-source model  (i.e. $E(b, h\rightarrow\infty)=W(0,\sqrt{4b})/2$)
up to the exponentially small  terms.
\\
 In the following section
an  approximate  expressions for the complementary
 range of parameters $b \cdot h^2<1$ will be derived.

\subsection{Power series: approximate expressions for $b\cdot h^2<1$}
\label{sec:MFLS}
 The integral  $E(b, h)$  (\ref{eqn:B3W})  can be approximated}  for small values of  the parameters $b$, $b\cdot h^2$ ($b<b\cdot h^2 \leq 1$) in the following way.\\
To   develop  approximate expressions for sufficiently small values of  $b\cdot h^2$ and  $h>1$,
the double  series ${\bb S}(b, h)$ from  Eq.(\ref{eqn:F}) is  decomposed on the sums over  the first two columns  in  $\{m, f\}$ index plane, complete diagonal and the remainder as follows
\begin{eqnarray}\label{eqn:Sbh2}
&& {\bb S}(b, h)= \sum\limits_{m=1}^\infty{m!(-1)^{m+1}\over [(m+1)!]^2} {1\over h^{2(m+1)}}\cdot\Gamma(m+3/2,0;b\cdot h^2)\cdot{1\over  2\sqrt{\pi}}\nonumber\\
&&+b\cdot\sum\limits_{m=1}^\infty{m!(-1)^{m+1}\over [(m+2)!]^2} {1\over h^{2(m+1)}}\cdot\Gamma(m+3/2,0;b\cdot h^2)\cdot{1\over  2\sqrt{\pi}} \\
&&-{1\over 2h\sqrt{b}}\big(I_0(\sqrt{4 b})-1\big)\cdot\exp(-2h\sqrt{b})\big(1+ {1\over 2h\sqrt{b}}\big)\nonumber\\
\nonumber\\
&&+R(q=2,p=2)\nonumber
\end{eqnarray}
 First, we obtain the $b$-series  using the linear and zero  approximations for the $\Gamma(m+3/2,0;b\cdot h^2)$ in the first   and second  terms in Eq.(\ref{eqn:Sbh2}), respectively, and then  we perform  exactly infinite summation over $m$.
Substituting  the  following asymptotic  estimate   for small values of  $b$   (Kilbas et al., 2010)
\begin{eqnarray}\label{eqn:kil}
&&\Gamma(\alpha,0;b)=\Gamma(\alpha) +\Gamma (\alpha)\cdot b/(1-\alpha)+\Gamma(-\alpha)\cdot b^{\alpha} +O(b^2),\\\nonumber
\\\nonumber
&&\alpha=m+3/2,     m=0, 1,..\nonumber
\end{eqnarray}
to the first  term  of the right-hand side of  Eq.(\ref{eqn:Sbh2}) and, then  identifying   the  power series expansions in $1/h^2$ (Prudnikov et al., 1983),
we arrive at
\begin{eqnarray}
\label{eqn:colD1}
&&-\sum\limits_{m=1}^\infty{m!(-1)^{m+1}\over [(m+1)!]^2} {1\over h^{2(m+1)}}\cdot\Gamma(m+3/2)\big(1+{b_H\over 3/2-m}\big)\cdot{1\over  2\sqrt{\pi}}\nonumber\\
&&=sinh^{-1}{h} - ln(2\cdot h) -{b\over2}\cdot{}_3F_2({1\over  2},1,1;2,2;-{1\over h^2})+{b\over 2} -{1\over 4h^2},
\end{eqnarray}
where  $\Gamma(m+1/2)=\Gamma(m+1/2,0;0)=\sqrt{\pi}{(2m)! / 2^{2m}}$ is the Gamma function  (Abramovitz   and  Stegun, 1965) and  ${}_3F_2({a_1,...,a_p},{b_1,...,b_q};z)$
is the generalized hypergeometric function  ${}_pF_q(a;b;z)$ (Gradshteyn  and   Ryzhik, 1980).
\\
Then, the summation identities (Prudnikov et al., 1983) can be also used to obtain the explicit form for the $b$ coefficient of the second term in Eq.(\ref{eqn:Sbh2})
\begin{eqnarray}
\label{eqn:colD2}
&& -b\cdot\sum\limits_{m=1}^\infty{m!(-1)^{m+1}\over [(m+2)!]^2} {1\over h^{2(m+1)}}\cdot\Gamma(m+3/2)\cdot{1\over  2\sqrt{\pi}}\nonumber\\
&&={-b\over 16 h^2}\cdot [1-{}_3F_2({3\over 2},1,1;3,3;-{1\over h^2})].
\end{eqnarray}
Because of the elementary form of the  generalized hypergeometric function  representations  for the given values of  arguments $a$ and $b$ in Eqs.(\ref{eqn:colD1}) and (\ref{eqn:colD2}) (Prudnikov et al., 1983),
we reduce the approximate expression for ${\bb S}(b, h)$ to the  simple form
\begin{eqnarray}\label{eqn:Sbh}
&&- {\bb S}(b, h)=\ln\big([1+\sqrt{1+1/h^2}]/2\big)\cdot\big(1+b\big)\nonumber\\
&&+{1\over 2h\sqrt{b}}\big(I_0(\sqrt{4 b})-1\big)\cdot\exp(-2h\sqrt{b})\big(1+ {1\over 2h\sqrt{b}}\big)\\
&&-{1\over 4h^2}-{b\over  2}\cdot\big(1+{1\over 8h^2}\big)+bh^2\cdot\big(-1+\sqrt{1/h^2+1}\big)\nonumber
\end{eqnarray}
where $R(q=2, p=2)$,   the remainder term  from Eq.(\ref{eqn:Rq}), which  is of the
order of $O(b)^{3/2}$, is neglected for small values of $b$ when $h>1$.\\
Using the first order expansion of  ${\bb S}(b,h)$ in $b$ (but not in  $h\sqrt{b}$) we arrive at new  approximate expression   for $E(b,h)$ from Eq.(\ref{eqn:F}), which can be used for small values $b\ll 1$
\begin{eqnarray}
\label{eqn:BB}
&&\tilde{E}(b,h)  =  W(0,\sqrt{4\emph{b}})/2-W(h\sqrt{4\emph{b}},0)I_0(\sqrt{4 b})\nonumber\\
&&+\ln\big([1+\sqrt{1+1/h^2}]/2\big)\cdot\big(1+b\big)\nonumber\\
&&+{\sqrt{b}\over 2h}\big(1+b/4\big)\cdot\exp(-2h\sqrt{b})\big(1+ {1\over 2h\sqrt{b}}\big)\\
&&-{1\over 4h^2}-{b\over  2}\cdot\big(1+{1\over 8h^2}\big)+b\cdot h^2\big(-1+\sqrt{1/h^2+1}\big)
\nonumber
\end{eqnarray}
In  the limiting case   of $b\rightarrow 0$    the   function  ${\bb S}(b, h)$ in Eq.(\ref{eqn:Sbh}) tends to
\begin{eqnarray}\label{eqn:BS}
&&\textbf{-}{\bb S}(h)=- ln(2\cdot h)+\sinh^{-1}{h}= \ln\big([1+\sqrt{1+1/h^2}]/2\big),
\end{eqnarray}
whereas  the   summation   of the \emph{well aquifer} functions in (\ref{eqn:BB}) is just the limit
\begin{eqnarray}
\label{eqn:lim}
&&\lim\limits_{z\rightarrow 0} [K_{0}(z)+Ei(-hz)I_0(z)]=ln(2\cdot h)
\end{eqnarray}
when $z(=\sqrt{4b})$ tends to zero (Abramovitz and  Stegun, 1965).  The right-hand side of (43) is regular at $z= 0$, because the singularities cancel each other.
 Therefore,  in this limiting case  the sum of  Eqs.(\ref{eqn:BS}) and (\ref{eqn:lim}) reduces to the steady-state Hantush's function
$E(b=0,h)=\sinh^{-1}{h}=M(0,h)/2$ (Trefry, 2005).
This comparison shows that  Eq.(\ref{eqn:BB}) is consistent with Eq.(\ref{eqn:M}), when $u\rightarrow 0$.  In the following, the developed approximate expressions will be validated numerically in the whole range of parameters.
\section{Numerical Validation}
\label{sec:PRE}
To verify the developed approximate expressions as means of calculating modified Hantush's function,
they are computed and  compared to exact values of the integral Eq.(\ref{eqn:A4}) in a wide range of parameters $b$ and $h$.
 The relative error  between them (Trefry, 2005)
\begin{eqnarray}
\label{eqn:N}
RE(b,h)=-Log_{10}(1- \tilde{E}/E)
\end{eqnarray}
calculates the number of  decimal places in which the  value of the approximation $\tilde{E}$ is equal to the one of the exact integral  $E$.  
 We used Mathematica  that numerically computes $E(b,h)$  estimates using standard quadrature rules (Wolfram, 2003). 
The results  of  such evaluation are   graphically represented  as  surfaces and contour lines
in  $b$ and $h$  shown in Fig.1, 2 and 3. \\
Fig.1  compares  relative errors  $RE(b, h)$ of two approximations (i.e. Eqs.(\ref{eqn:B8}) and (\ref{eqn:BB}))  for $b\leq 0.1$, which are  represented as surfaces intersecting  on the transition line  $b \cdot h^2=1$ in Fig.1a.
Fig.1b   shows it as red line   separating  left- and  right-hand segments
 of validity  for Eq.(\ref{eqn:BB})   and  Eq.(\ref{eqn:B8})}, $b \cdot h^2 < 1$  and $b \cdot h^2 > 1$, respectively.
Fig.1 (a and b)  shows  the regions of accuracy  of  more than $5$ to $8$ decimal places  to the left from the transition line and  from $4$   up to   $7$ to the right from the line.
 One might also note in Fig. 1c that  the accuracy  for two approximate expressions  increases (up to   $11$ decimal places) with $h$
at the extended interval of parameter values.  
This precision is sufficient for practical evaluation of the integral within the selected  range of the $\{b, h\}$ parameter space.
\\
Fig.2 illustrates  for  $b \cdot h^2>1$  how the distribution   of accuracy for the  approximation  in  Eq.(\ref{eqn:AB9}), which includes the series truncated to only one term,
differs  from that  in Eq.(\ref{eqn:B8}), which includes the series truncated to three terms, shown in Fig.1.
The  approximate expressions (\ref{eqn:B8})  and   (\ref{eqn:AB9}) provide  maximum  precision with  $RE$  values of $6$  and   $7$,  as shown in Fig.1b and 2b.
Figures 1b and 2b also indicate that the relative errors of the  approximations (\ref{eqn:B8}) and (\ref{eqn:AB9}) on the transition line of $b\cdot h^2=1 $  turn out to be close  to ones calculated from  approximate series expansion (\ref{eqn:BB}) so that the  contour $RE(b, h)$ lines are nearly continuous in Fig.1a and 2a.
One might also observe in Fig. 2(a) and (b) that the asymptotic approximation (\ref{eqn:AB9}) is   more accurate than
the one  proposed in Eq.(\ref{eqn:B8}) at the  narrow domain to the right from the  transition curve.
\\
Fig.3 compares accuracy of the  approximation  (\ref{eqn:B8})  to that  calculated  with Eq.(\ref{eqn:AB9}) at the extended range of parameter  $b$ values  from $0.1$ to $1$   (shown in Fig.3a and 3b).
 While both approximations are  comparable in accuracy in the interval of $b<0.1$, where it is sufficient  to  use only one term from the infinite series (\ref{eqn:B4}) in Eq.(\ref{eqn:AB9}),
the approximate  expression (\ref{eqn:B8}), which includes the series truncated to three terms,  provides accurate estimation on the larger region than that given in Eq.(\ref{eqn:AB9}).  Thus, the asymptotic expansions in Eqs.(\ref{eqn:B8}), (\ref{eqn:AB9}) and   (\ref{eqn:BB}) are
justified  by comparing  with the exact integral $E(b,h)$  (\ref{eqn:A4}).\\
 Note that the  computation of  $\tilde{E}$  estimates is  from 70 to 20 times faster than the numerical integration of   $E(b,h)$ over their parametric space shown in Fig.1, 2 and 3. Therefore, use of the analytical  accurate approximations  of the integral results in a noticeable decrease of the CPU time when compared to the numerical integration.\\
To summarize, Fig. 1, 2  and 3  show a good agreement between  the exact integral and its  approximations  at the domains spanned by parameters: $b<b\cdot h^2<1$ and $b\leq 1\leq b\cdot h^2$.
For the small values of  parameter  $b \cdot h^2$  ($b \cdot h^2<1$)
 Eq.(\ref{eqn:BB}) is proposed to use  for computing modified Hantush's  function.
  For the large values of  the parameter,  $b \cdot h^2\geq1$, the use of Eq.(\ref{eqn:B8}) for $b<1$ or
Eq.(\ref{eqn:AB9}) for $b<0.1$  is proposed.


\section{Conclusions}\label{sec:con}
Results have been presented of a study of  modified Hantush's function that arises in modeling  of heat pumping from subsurface with groundwater flow  under steady-state conditions.
Exact expression  for  this integral function has been obtained in the form  that includes  special  functions and  rapidly convergent infinite sum over   generalized incomplete gamma functions.   As a by-product, new closed-form for subsidiary integral has been derived.\\
On this basis we have  provided new approximate expressions
for calculating  modified Hantush's function $E(b,h)$
over  a wide   range of parameter values.
 The  relative   error between  the  results of numerical integration of $E(b,h)$ and approximate expressions does not exceed   $0.01\%$. These  low-order  asymptotic expansions of high accuracy might be useful for practical applications.
\\Analytical expressions have been obtained for the asymptotic behavior of the $E(b, h)$ function in the two limiting cases of $b\rightarrow 0$ and $h\rightarrow \infty$ when  it recovers  classical Hantush's  functions $M(0,h)$  and  $W(0,\sqrt{4b})$.
The corrections to them  that depend on  the both parameters have been  provided in elementary functions.

\medskip
\noindent {\bf Acknowledgments}.  This paper was partially supported by the Spanish Ministry of Science and Innovation  under the  project  ENE2012-38633-C03-03.
\medskip
\noindent
\medskip
\newpage
\noindent
\medskip
\medskip
\noindent
\medskip
\label{sec:refs}
{\bibliographystyle{unsrt}
\newpage
 \newpage
\bf{FIGURE CAPT   IONS}
\begin{itemize}
\item \bf{Figure 1}
{\it  Approximate functions compared to exact ones calculated from Eq.(\ref{eqn:A4}).  Relative error (RE) less than $10^{-4}$ as function of $b$ and $h$  calculated from Eq.(\ref{eqn:N}) at the domain  $b\in [0,0.1]$,  $h \in [4, 10]$.
There are two segments separated by  the transition  line  $h=1/\sqrt{b}$:
(a) The intersection of the approximations from  Eqs.(\ref{eqn:B8}) and (\ref{eqn:BB});
(b) The contour region to the left (right) from the $b=1/h^2$ curve (red solid line) calculated from Eq.(\ref{eqn:BB}) (Eq.(\ref{eqn:B8}));
(c) The contour region to the left (right) from the $b=1/h^2$  curve (red solid line)  evaluated from Eq.(\ref{eqn:BB}) (Eq.(\ref{eqn:B8}))
over the extended interval of the parameter $h \in [4,100]$.}
\item \bf{Figure 2}
{\it  Relative error (RE)  as function of $b$ and $h$ at the segment $b>1/h^2$ where approximate function evaluated from Eq.(\ref{eqn:AB9}):
(a) The intersection of the approximate surfaces calculated from  Eqs.(\ref{eqn:AB9}) and (\ref{eqn:BB});
(b) The contour region to the left (right) from the $b=1/h^2$  curve (red line) evaluated from Eq.(\ref{eqn:BB}) (Eq.(\ref{eqn:AB9})).}
\item \bf{Figure 3}
{\it Comparison between approximations   in  Eq.(\ref{eqn:B8})(a)  and
(\ref{eqn:AB9})(b) at the domain $b\in [0, 1]$,  $h \in [4, 10]$. Relative error (RE) in the range  from $10^{-4}$ to $10^{-10}$ as function of $b$  and $h$:
(a)} {\it The contour region  to the  right  from the $b=1/h^2$  curve (red line) calculated from  Eq.(\ref{eqn:B8}); (b) The contour region  to the  right  from the $b=1/h^2$  curve (red line) calculated from  Eq.(\ref{eqn:AB9}).}
\end{itemize}
\newpage

\begin{center}
\textbf{FIGURE 1}
\end{center}
\begin{figure}[b!]
\begin{center}
\small
\begin{tabular}{ccccc}
\\
\hline
(a)
\\
\IG[scale=0.65]{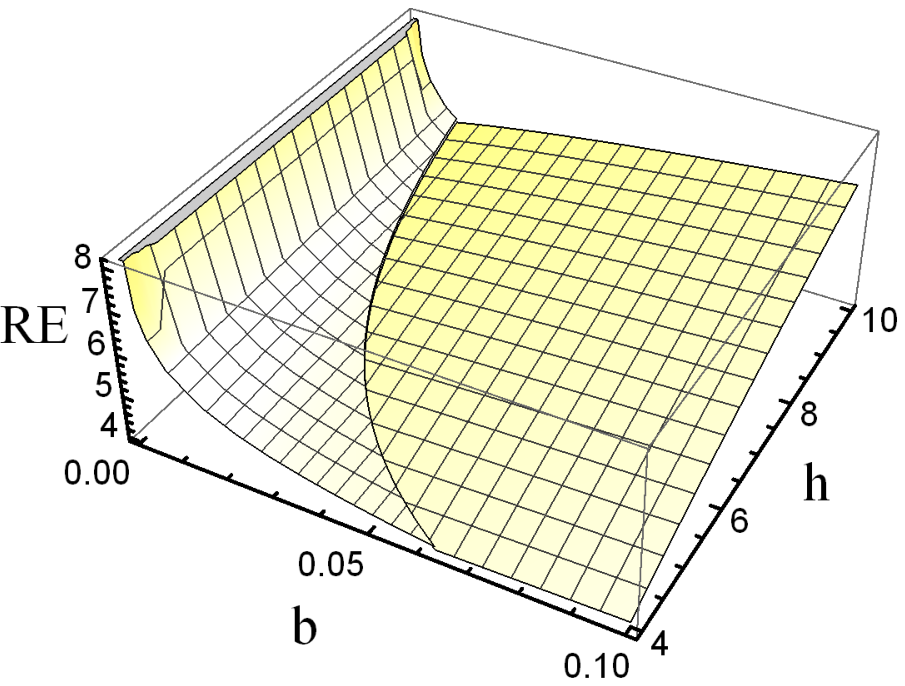}
\\
(b)
\\
\IG[scale=0.6]{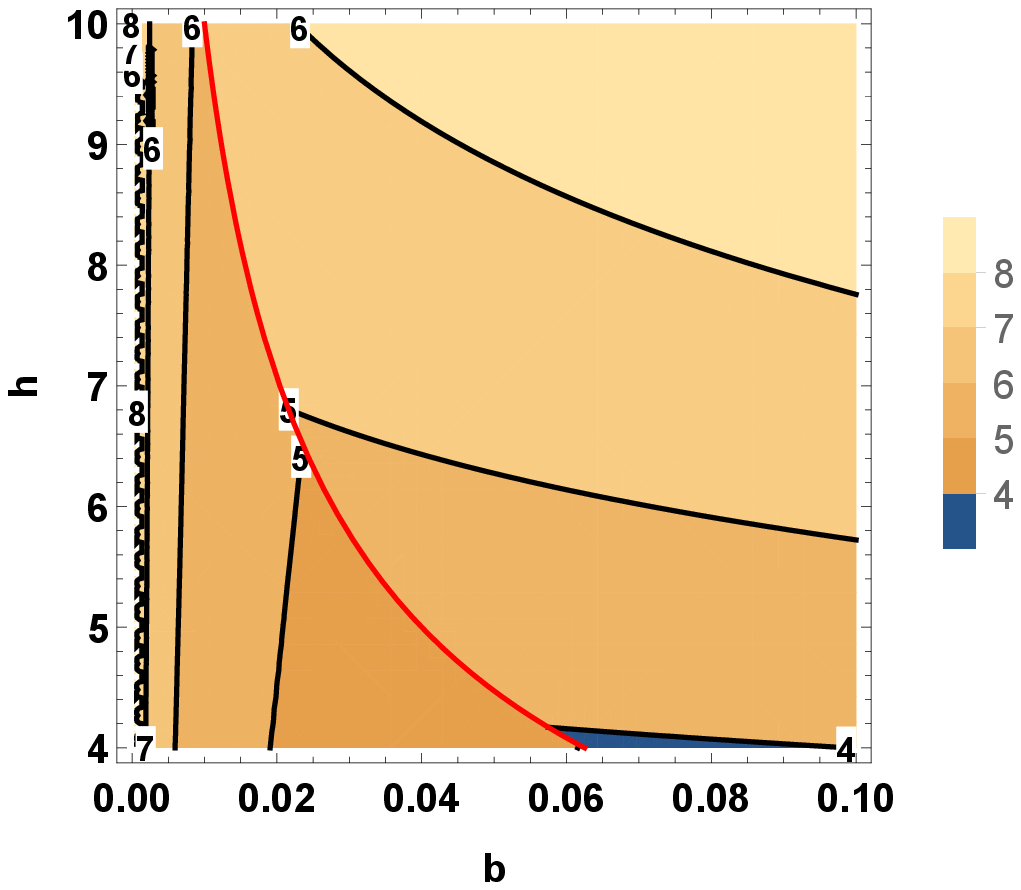}
\\
(c)
\\
\IG[scale=0.6]{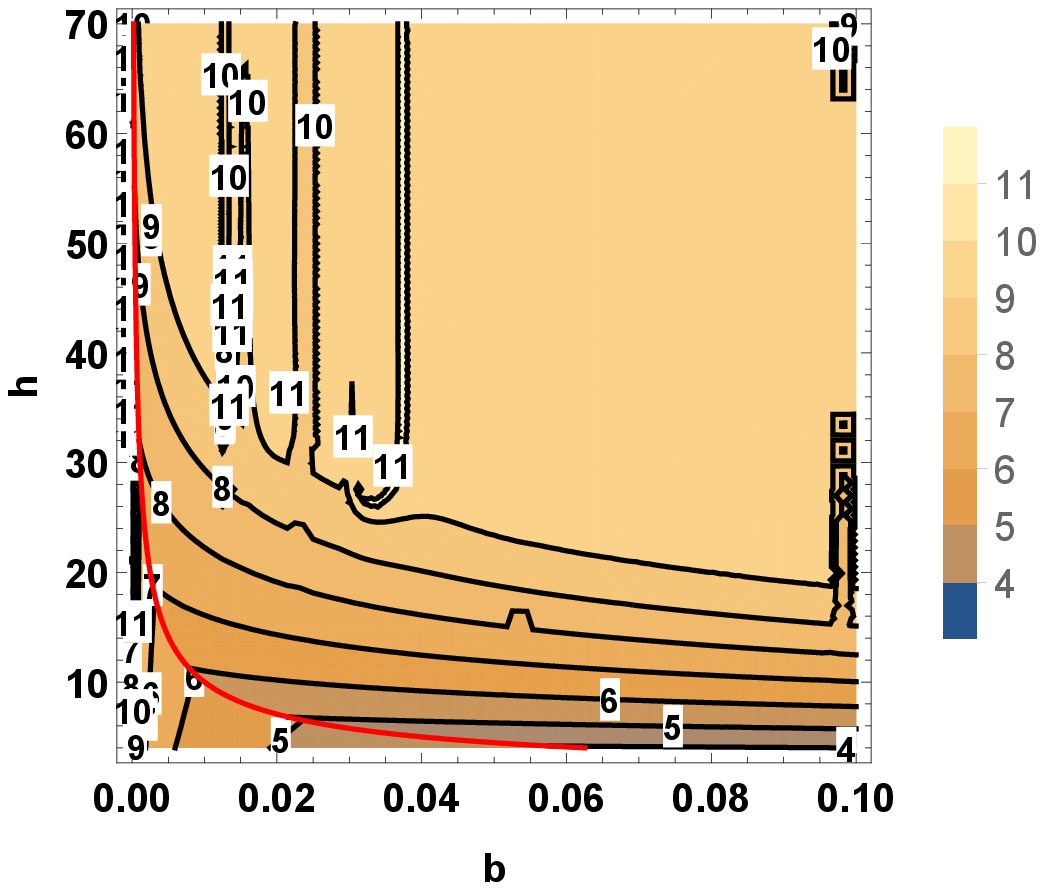}
\\
\hline
\end{tabular}
\end{center}
\label{figure1}
\end{figure}
\newpage
\begin{center}
\textbf{FIGURE 2}
\end{center}
\begin{figure}[b!]
\begin{center}
\small
\begin{tabular}{ccccc}
\\
\hline
(a)
\\
\IG[scale=0.95]{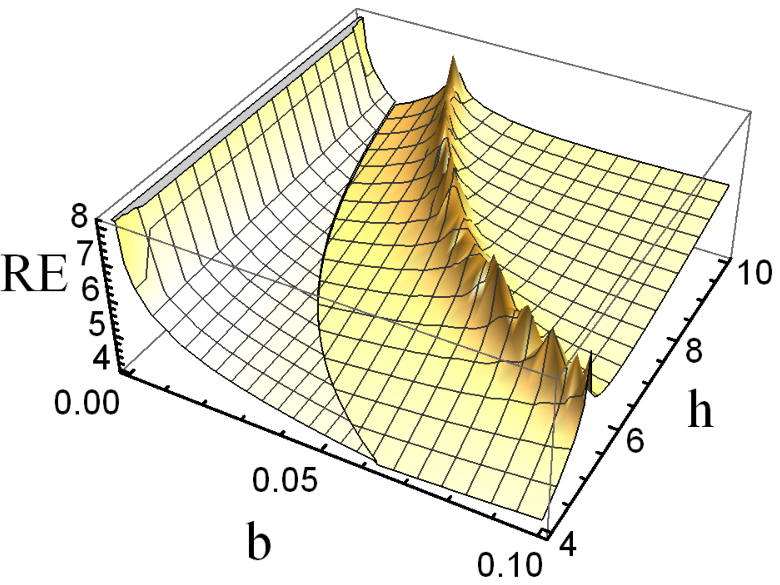}
\\
(b)
\\
\hspace{0.4cm}
\IG[scale=0.95]{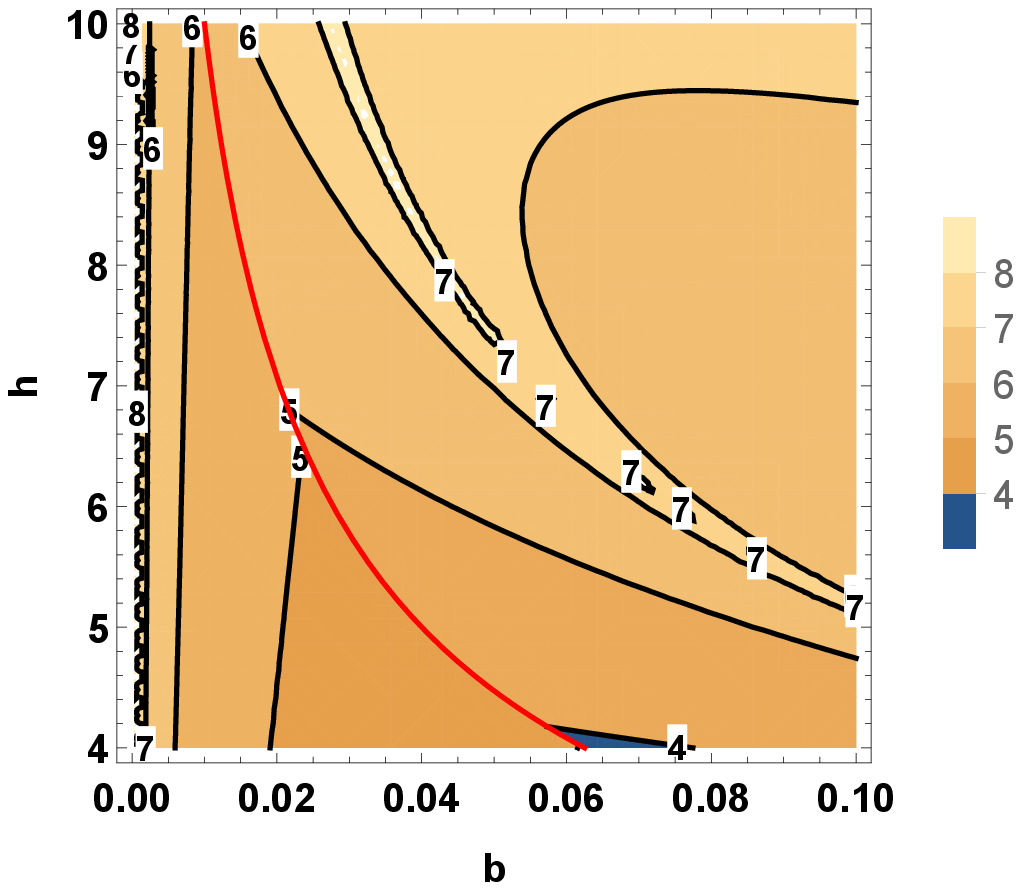}
\\
\hline
\end{tabular}
\end{center}
\label{figure4}
\end{figure}
\newpage
\begin{center}
\textbf{FIGURE 3}
\end{center}
\begin{figure}[b!]
\begin{center}
\small
\begin{tabular}{ccccc}
\\
\hline
\\
(a)
\\
\IG[scale=0.95]{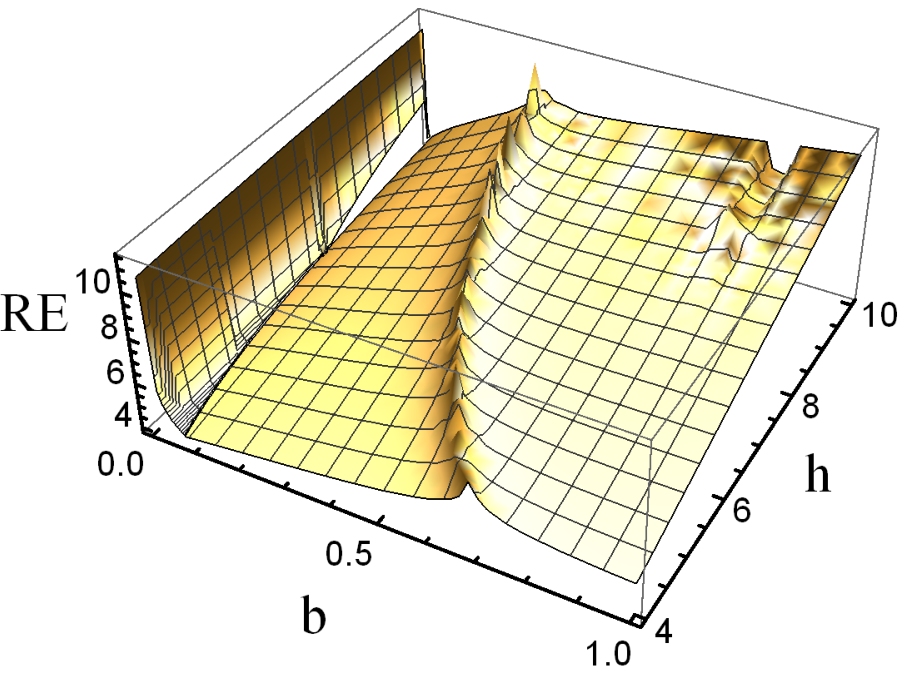}
\hspace{0.5cm}
\IG[scale=0.7]{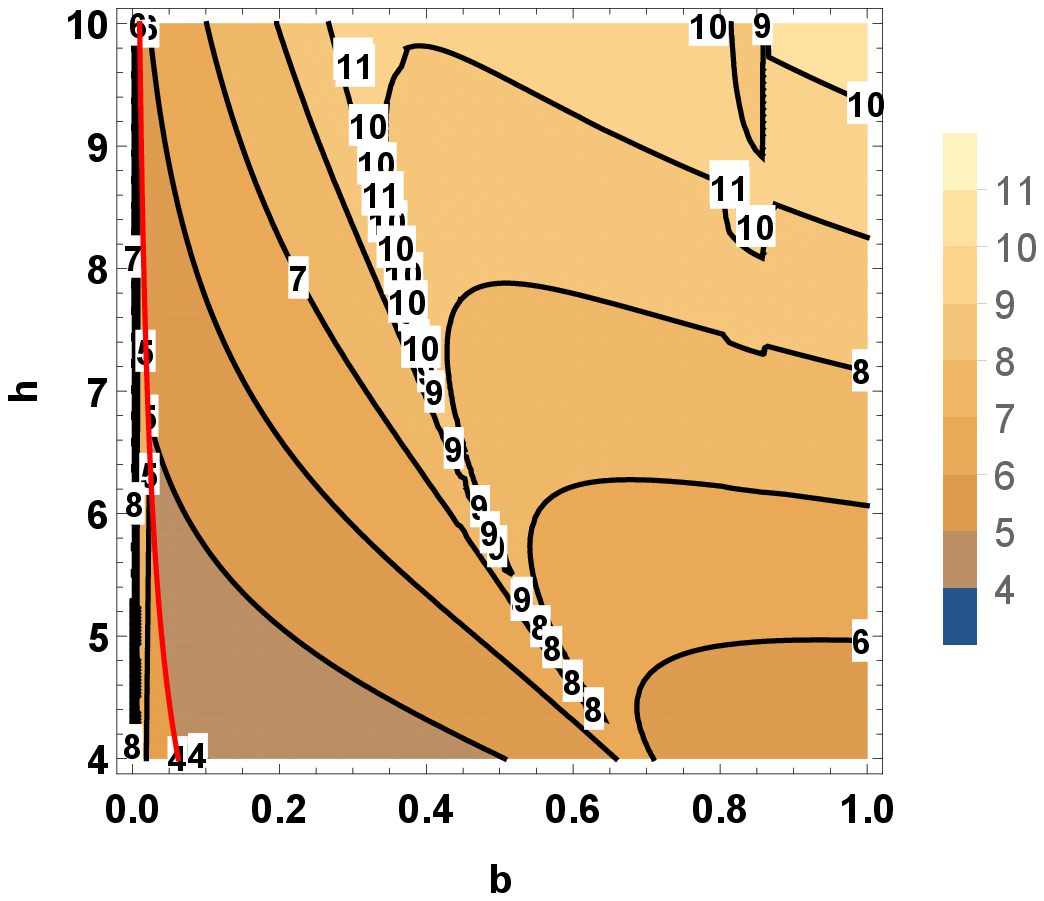}
\\
\hline
\\
(b)
\\
\IG[scale=0.95]{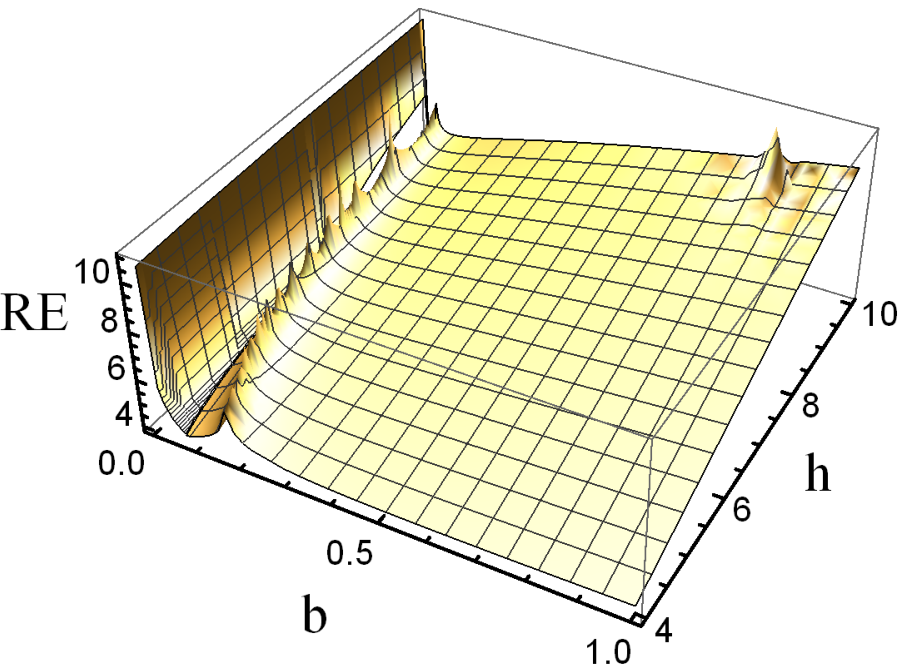}
\hspace{0.5cm}
\IG[scale=0.7]{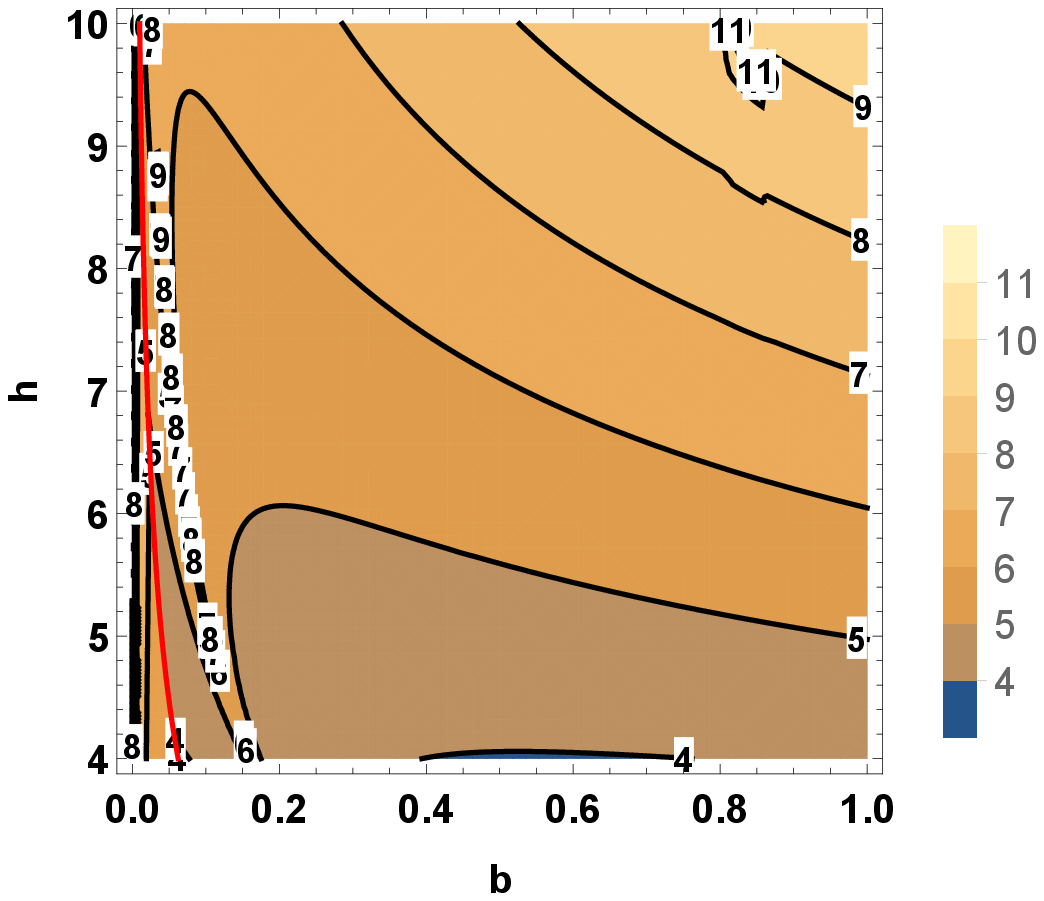}
\\
\hline
\end{tabular}
\end{center}
\label{results2}
\end{figure}
\newpage

\label{sec:nomencl}
\begin{table}[b!]
\centering
\small
\begin{tabular}
{|p{4.7cm} p{11.cm}|}
\hline
     {}    &  \textbf{  Nomenclature  }
\\
$b$                             &  parameter in Eq.(\ref{eqn:A4}) (-)   
\\
$b_H(=bh^2)$                    &  parameter in Eq.(\ref{eqn:AG})  (-)
\\
$\tilde{E}(b, h)$               & approximate  modified Hantush function (-1)
\\
$E(b, h)$                       & modified Hantush function, Eq.(\ref{eqn:A4}),  (-)
\\
$Ei(u)$                         & exponential integral  (-)
\\
$(erfc)erf$                     & (complementary) error function (-)%
\\
$K_0(I_0)$                      & zero-order modified Bessel function of the second (first) kind (-)
\\
$K_{m+1/2}$                     & half-integer order  modified Bessel function of the third kind (-)
\\
$h$                             & dimensionless line-source length (-)
\\
$M(u, h)$                       & Hantush's integral, Eq.(\ref{eqn:M}),  (-)
\\
$t$                             &   time ($s$)
\\
$W\big(u, \beta=\sqrt{4 b}\big)$  &  Hantush-Jacob well function for leaky aquifers,    Eq.(\ref{eqn:W}), (-)
\\
$W(u)(=-Ei(-u))$                        & well function for non-leaky aquifers  (-)
\\
$W(0,\beta)(=2K_0(\beta))$         & steady-state well function  for leaky aquifers (-)
\\
{ \emph{Greek letters}  }      & {\textbf{ }}
\\
$\gamma$                        & Euler's constant (-)
\\
$\Gamma(0,u;{\beta^2\over 4})\big(=W(u,\beta)\big)$  & generalized  incomplete  gamma function \cite{C&Z94}(-)
\\
\hline
\end{tabular}
\label{tab:ps}
\end{table}
\end{document}